# snapshot CEST++ : the next snapshot CEST for fast whole-brain APTw imaging at 3T

Patrick Liebig,[1] Maria Sedykh[2], Kai Herz[3,4], Moritz S. Fabian[2], Angelika Mennecke[2], Simon Weinmüller[2], Manuel Schmidt[2], Arnd Dörfler[2], and Moritz Zaiss[2, 3]*

[1] Siemens Healthcare GmbH, Erlangen, Germany

[2] Institute of Neuroradiology, Universitätsklinikum Erlangen, Friedrich-Alexander Universität Erlangen Nürnberg, Germany

[3] Magnetic Resonance Center, Max-Planck-Institute for Biological Cybernetics, Tübingen, Germany

[4] Department of Biomedical Magnetic Resonance, University of Tübingen, Tübingen, Germany
*moritz.zaiss@uk-erlangen.de

# Introduction

Amide proton transfer weighted imaging is a metabolic MRI technique based on chemical exchange saturation transfer (CEST) with the ability of tumor detection and subtyping[1]. CEST MRI suffers from long preparation times and consequently long acquisition times as well as the interplay of the acquisition onto the preparation state itself. To reduce acquisition time, snapshot CEST has been suggested[2]. Snapshot CEST acquires the entire 3D volume after one preparation block. This reduces the acquisition time to an absolute minimum. Furthermore, snapshot CEST completely disentangles the preparation from the readout block, giving the freedom to optimally design the preparation block for the contrast desired.

Snapshot GRE CEST is currently limited to a smaller volume of 12-18 slices. The reason for the lower coverage of the GRE acquisition is the limited number of k-space lines that can be acquired after one preparation phase. Previously we showed that roughly 800 k-space lines can be acquired with minimal loss of contrast and not too high increase of blurring[1]. GRE-CEST on the other hand has the advantage of avoiding the typical artefacts that come with EPI typed acquisitions, like B0 induced distortions[3,4]. It already imposes a challenge for brain imaging, whereas for body or even msk imaging it becomes a severe obstacle. To overcome the lower coverage of the snapshot GRE CEST we want to present an extended version utilizing compressed sensing[5]. The rational is that with the limit of 800 k-space lines a larger volume can be acquired, when these lines are sampled incoherently, and the data is reconstructed iteratively. We show herein that this allows to extend the volume and resolution of the slab selective snapshot GRE CEST to whole brain acquisition.

As one of the most used CEST contrasts, we focus herein on application of the CEST sequence for APTw-imaging, which is defined as $MTR_{asym}$(3.5 ppm, $B_{1rms}$ = 2 µT). The article is organized in two parts: in part I we show improvements and adjustments for high-quality APTw imaging using the conventional snapshot GRE CEST approach. This part serves as reference for conventional snapshot APTw imaging sequences currently in use, as well as for part II.

In part II, we show that these APTw images can be reproduced with the compressed sensing version of the snapshot CEST sequence (cs-snapshot-CEST in the following), with increased volume coverage.

# Material and Methods

A prototype MR sequence was applied in healthy volunteers and one patient under approval of our local ethics committee. Scanning was performed on a 3T whole-body MRI system (MAGNETOM Prisma, Siemens Healthcare, Erlangen, Germany) and a 64 Channel receive coil.

**standardized CEST preparation based on the pulseq-CEST library**

For the CEST saturation block, parameters optimized according to Herz et al. were used[6]. The exact definitions of the preparation modules were defined using Pulseq-CEST (https://pulseq-cest.github.io/) and are available in the Pulseq-CEST library (https://github.com/kherz/pulseq-cest-library) by their identifiers:

**APTw_3T_001:** pulse train of 36 sinc-gauss pulses of tp=50ms, td=5ms, DC=91%, $T_{prep}$=1.975 s, $B_{1rms}$ =2 µT

**APTw_3T_002:** pulse train of 20 sinc-gauss pulses of tp=50ms, td=50ms, DC=50%, $T_{prep}$=1.950 s, $B_{1rms}$ =2 µT

**APTw_3T_003:** pulse train of 8 block pulses of tp=100ms, td=50ms, DC=95%, $T_{prep}$=0.834 s, $B_{1rms}$ =2 µT

APTw_3T_001 and 003 also correspond to the recent recommendation of the APTw consensus white paper[7]. One crusher gradient was applied after the pulse trains to spoil spurious transverse magnetization. Saturation and readout were repeated at 30 off-resonance frequencies in the range between -5 ppm and 5 ppm, and at -300 ppm for an unsaturated reference image. A recovery time of 3.5 s was placed before each pulse train saturation and readout resulted in an acquisition time per offset of TA = 6.1 s. For 30 frequency offsets and a 12-second recovery time before the first CEST module, this yielded a total scan time of approximately 3 min for one APTw snapshot CEST acquisition.

**snapshot CEST sequence and reconstruction**

The conventional 3D snapshot-CEST sequence[2] was composed of the described preparation blocks and a 3D RF- and gradient-spoiled gradient-echo readout with Cartesian centric rectangular spiral reordering (elongation factor E = 0.5). Initial readout parameters were FOV = 220 x 180 x 60 mm$^3$ and matrix size 128 x 104 x 12 for 1.7 x 1.7 x 5 mm$^3$ resolution, TE = 2 ms, TR = 4 ms, bandwidth = 700 Hz/pixel and flip angle FA = 7°. Different matrix sizes and FA were studied. Parallel imaging (GRAPPA[8]) and omitting the corners of the k-space (elliptical scanning), the total number of k-space lines was 666 with the standard settings, leading to a readout time of $t_{RO}$=2.7 s. The reconstruction for this sequence is a conventional GRAPPA reconstruction with adaptive coil combination; only magnitude images were generated for further evaluation. This version corresponds to the MPI04 sequence; complete protocols can be found at https://github.com/cest-sources/cestineers .

**cs-snapshot CEST sequence and reconstruction**

The compressed sensing 3D snapshot-CEST sequence is an adaption of[2], but based on the volumetric interpolated breath-hold examination (VIBE) sequence[9]. The same described preparation blocks were followed by the adapted VIBE sequence. The 3D VIBE readout is a RF- and gradient-spoiled gradient-echo readout with a flexible Cartesian centric spiral reordering, ordering the acquired lines on an ellipsis in the y-z-direction, with different sampling patterns as described in the reconstruction paragraph below. Different matrix sizes and acceleration factors were studied with the constraint to keep the number of acquired k-space lines the same between all settings.

The reconstruction for this sequence is a L1-regularized, fully coupled CS reconstruction based upon a sparse incoherent spatio-temporal Poisson disk undersampling, as described in[10]. The reconstruction itself was running online on the scanner hardware. Four different schemes were tested: 1) the acceleration factor was kept the same as in the standard parallel imaging case to compare how the new sequence and reconstruction will impact the APT contrast. 2) The acceleration factor was increased to 11.04 to enable a whole brain 2 mm isotropic acquisition. 3) The acceleration factor was further increased to 14 to enable a 1.8 mm isotropic whole brain acquisition, as achieved in [4] 4) The undersampling mask was varied for the 1.8 mm isotropic protocol and regularization along the offset dimension was introduced. Required coil sensitivities were estimated using ESPIRiT[11]. Reconstruction times are 3-5 minutes depending on the matrix size as well as the number of offsets.

For all settings, only magnitude images were generated for further evaluation. The protocol parameters can be seen in table 1.

| Protocol | Slab selective | 2.0 whole brain | 1.8 whole brain |
| --- | --- | --- | --- |
| CEST preparation | 36 pulses, 50 ms pulse duration, 5 ms interpulse delay, B1 2.02 uT (Continuous wave power equivalent), Recovery time 1 s. | | |
| Resolution (RO x PE x PAR) [mm$^3$] | 2 x 2 x 5 | 2 x 2 x 2 | 1.8 x 1.8 x 1.8 |
| FOV (RO x PE x PAR) [mm$^3$] | 220 x 180 x 60 | 220 x 180 x 144 | 220 x 182 x 144 |
| CS acceleration | 2 | 8.66 | 10.98 |
| Number of lines per Shot | 666 | 666 | 666 |
| TR/TE [ms] | 4.11/2.08 | 3.07/1.34 | 3.2/1.39 |
| Flip angle [°] | 7 | 7 | 7 |
| Measurement time | 1min 23s | 1min 26s | 1min 28s |

*Table 1: Protocol Parameters*

**B0 mapping**

Two different methods were used for $B_0$ mapping, the WASABI mapping approach[12], and a GRE phase mapping approach[13].

**post-processing of all APTw-CEST data**

Both sequences generated magnitude data, which were further processed in a pixel-wise manner. B0 correction using the external $B_0$ maps was realized by a smoothing-spline interpolation (smoothing parameter 0.95). The subsequent pixel-wise asymmetry analysis was evaluated at 3.5 ppm, yielding the map MTR$_{asym}$(3.5 ppm) which is displayed using a rainbow color map.

**reproducibility evaluation using the standard deviation between measurements**

Standard deviation maps were generated by acquiring the APT contrast (using the below explained 11 offset scheme) 9 times, once for the conventional snapshot CEST with the limited slab excitation and once for the cs-snapshot CEST with a whole brain 2 mm isotropic protocol. Standard deviation maps were calculated by first registering and reslicing all reconstructed APTw maps onto the limited slab of the slab selective acquisition (first offset). On these CEST maps a pixel-by-pixel standard deviation was computed. This gives a reliable measure of how much the signal is fluctuating between measurements.

Results

# Part I – snapshot GRE CEST for APTw imaging

The first observation when comparing APTw images to the literature is that published results of healthy volunteers show a relatively homogenous green image[1], as shown in Figure 1d. Whereas our obtained images showed quite some gray-white matter contrast (Figure 1a-c). We highlight this here, as many researchers trying to reproduce APTw images think at this point their sequence or protocol is wrong and give up. However, while it is important to use the exact same preparation, also, the visualization is important. All the images in Figure 1 show the exact same APTw image. To get the previously published results it is necessary to use the same colormap, which in most of the published work is the rainbow colormap of IDL. This colormap has the feature of a large greenish range. With the same data, the image using this colormap and the right windowing between -5% and +5% of $MTR_{asym}$ yields the homogeneous green APTw image (Figure 1d). This is the colormap we will use here in the following; and the hallmark of APTw imaging to generate homogeneously green maps in healthy subjects is also our goal here. Still, $MTR_{asym}$ maps are not as smooth as they appear, but APTw imaging has a certain GM/WM contrast as can be seen in Figure 1a-c.

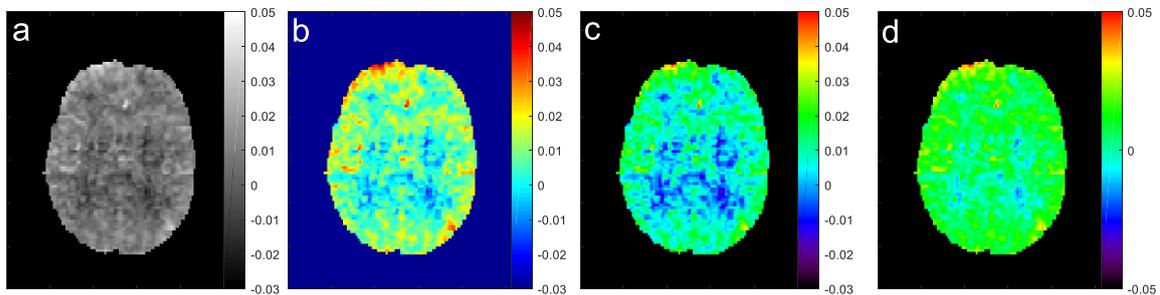

*Figure 1: APTw imaging (following APTw_3T_002) in a healthy volunteer by $MTR_{asym}$(3.5 ppm, $B_{1rms}$ =2 µT). (a-d) are different visualization of the exact same image just with different colormaps and/or windowing. a: grey[-3 5]%, b:jet [-3 5]%, c:rainbow [-3 5]%, d:rainbow [-5 5]%. (d) shows the homogeneous green background known from previous published work and using the IDL RAINBOW color map that can be found at https://github.com/planetarymike/idl-colorbars-python/blob/master/idl_colorbars/IDL_rgb_values/013_RAINBOW.dat*

Next, we optimized the snapshot CEST readout in terms of SNR. The snapshot CEST sequence yields still noisy maps when using the protocol as suggested for multi-pool CEST by Deshmane et al.[14]. In multi-pool acquisitions, many offsets are acquired, and the subsequent Lorentzian fitting forms an averaging and increases CNR. In APTw CEST less offsets are acquired leading to generally noisier images using the same readout. To increase the SNR we can chose higher readout flip angles (Figure 2b) or a larger voxel size (Figure 2c). As high flip angles cause stronger blurring[2], we followed the protocol of dynamic glucose-enhanced imaging at 3T of Herz et al.[15] and increased the voxel size to 2x2x5 mm³ and minimally increased the flip angle to 7°. As seen in Figure 2c this yields a smooth image with reasonable CNR.

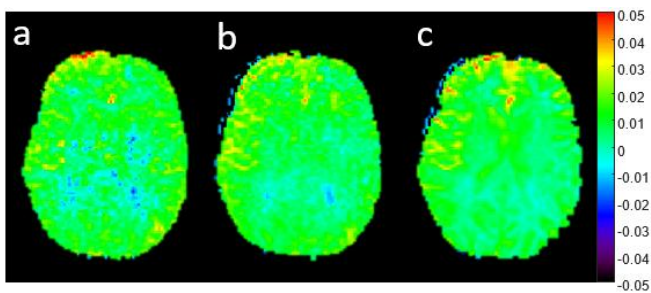

*Figure 2: Comparison of APTw images (APTw_3T_002) upon change of flip angle and voxel size. (a) 1.7x1.7x5mm, FA=5°, (b) 1.7x1.7x5mm, FA=9°, and (c) 2x2x5 mm³, FA=7°. The relative SNR of the readout (relative to a) is: 1.0   1.6   1.94*

Having sufficient SNR of the readout, we compared different APTw presaturation schemes in use, all with $B_{1rms}$ = 2 µT but with different rf duty cycles and saturation duration. Figure 3 reveals that the highest duty-cycle of 90% and the longest saturation duration of 2 s yielded best CNR in both healthy subjects and a tumor patient, showing the typical hyper-intensity in the tumor area.

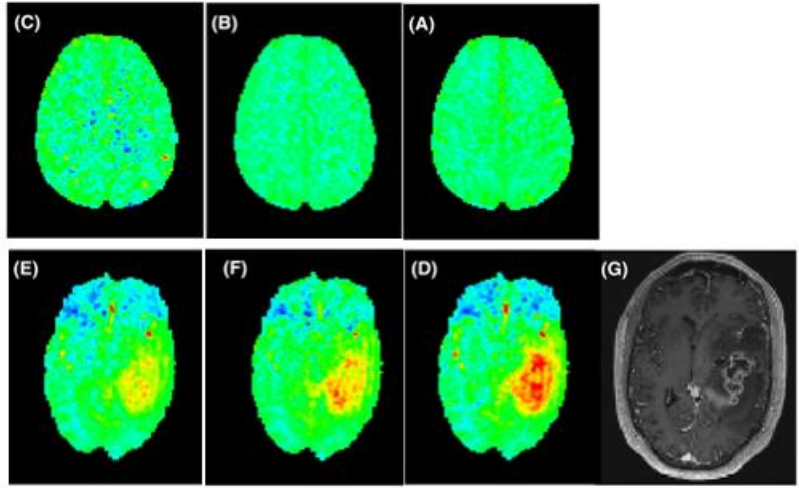

*Figure 3: APTw images in a healthy subject (a-c) and a brain tumor patient (d-f) using different preparation modules as defined in [https://github.com/kherz/pulseq-cest-library](https://github.com/kherz/pulseq-cest-library): (a,d) APT_001 with 90% duty-cycle and 2 s of saturation duration, (b,e) APT_002 with 50% duty-cycle and 2s saturation duration, (c,f) APT_003 with 95% duty-cycle and 834 ms of saturation duration. APTw_3T_001 shows strongest contrast in the tumor area, as outlined by the ce-T1 image (g). Figure reproduced from[6]*

Based on Figure 2 and 3 the our choice for the APTw snapshot CEST is the presaturation APTw_001 together with the slab selective snapshot GRE readout at 2x2x5mm, FA=7° as given in table 1.

For this snapshot CEST version we retrospectively reduce the offsets for APTw imaging, and we can confirm previous findings that few offsets sampled around ±3.5 ppm are sufficient to obtain APTw contrast[7,16] also with the snapshot CEST approach. As less as 7 measurements (M0 , ±3, ±3.5 ±4) are possible to still see the original contrast in healthy subjects and in a tumor patient with decent CNR (Figure 4) and unchanged robustness against $B_0$ inhomogeneities. We choose herein the more conservative compromise of 11 measurements (M0 , ±3, ±3.5, ±3.5 , ±3.5 ±4) with a measurement time of approximately 66 s, which still beats existing TSE-based APTw imaging sequences that typically take 5-10 minutes with comparable coverage.

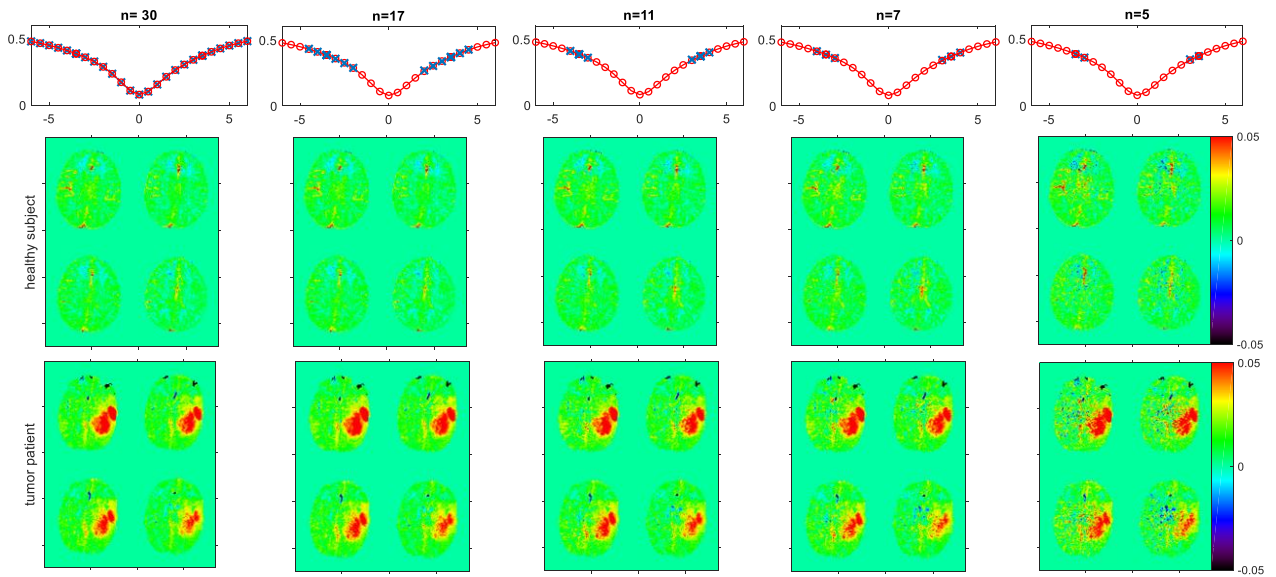

*Figure 4: Possibility of reduction of offset sampling for snapshot APTw imaging. As few as n=7 is possible with still reasonable CNR. The measurement number n includes the normalization scan which is not shown in the top row.*

# Part II – cs-snapshot CEST for APTw imaging

With this optimized APTw imaging using the conventional snapshot CEST approach, we have established a reference for validation of the APTw-cs-snapshot-CEST sequence.

The cs-snapshot CEST was set up using the same pre-saturation module APTw_001 followed by the cs-GRE readout. Figure 5 shows the direct comparison of the conventional slab-selective snapshot CEST and the slab-selective cs-snapshot CEST in the same healthy subject. The slab-selective cs-snapshot CEST in Figure 5b has additional slices and thus whole-brain coverage. In Figure 5c, the number of slices was doubled and the slice thickness halved, leading to same coverage but 2x2x2.5mm resolution. This is almost isotropic resolution, which we aim for in the final protocol.

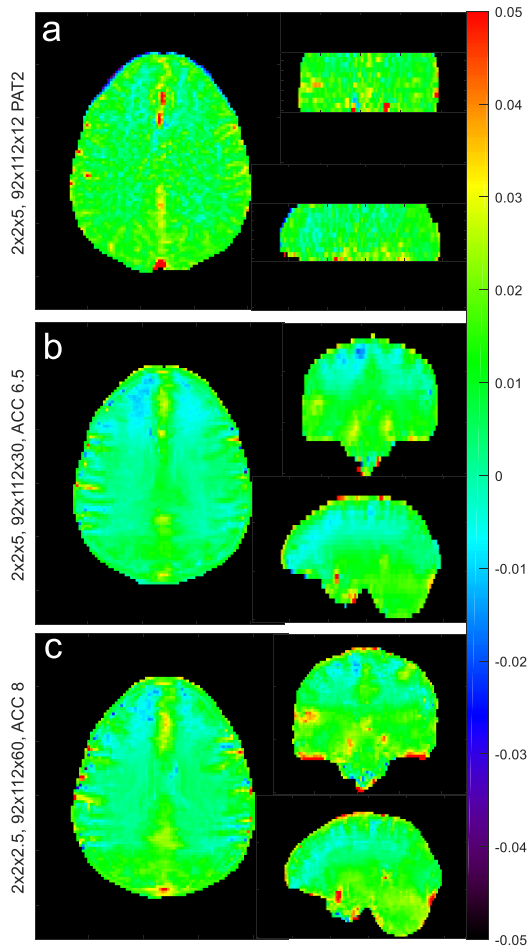

*Figure 5: The coverage of the conventional snapshot CEST (a) which is limited to a slab (here 12 5mm slices) is extended by the compressed sensing snapshot CEST approach increasing the number of slices to 30 5mm slices for whole-brain coverage as seen in transversal, sagittal and coronal view (b). Further increase of the acceleration and undersampling allows for 60 slices of 2.5 mm thickness yielding the same homogeneous contrast in the healthy brain, validating the cs-snapshot CEST approach.*

Figure 6 shows the comparison between the slab-selective protocol and a 2 mm isotropic whole brain protocol. No obvious difference can be seen. However, the standard deviation, as seen in the maps at the bottom row of figure 6, is about 16% higher for the whole brain protocol (Table 1). With Figures 5 and 6 we have validated that the cs-snaphsot CEST yields comparable contrast as the conventional snapshot CEST protocol, without much loss of standard deviation despite the increased coverage and resolution. As the number of acquired k-space lines is also constant, the scan time of the isotropic whole brain sequence is not significantly prolonged (compare table 1).

In a last step, we want to push the cs-snaphsot approach to its limit. By further increasing the undersampling factor per offset and making use of variable undersampling along the offset domain even higher resolutions are possible. Figure 7 shows the reconstructed APT maps of further increased resolution, showing that a 1.8 mm isotropic whole brain acquisition is possible without loss of acquisition time and minimal loss of SNR.

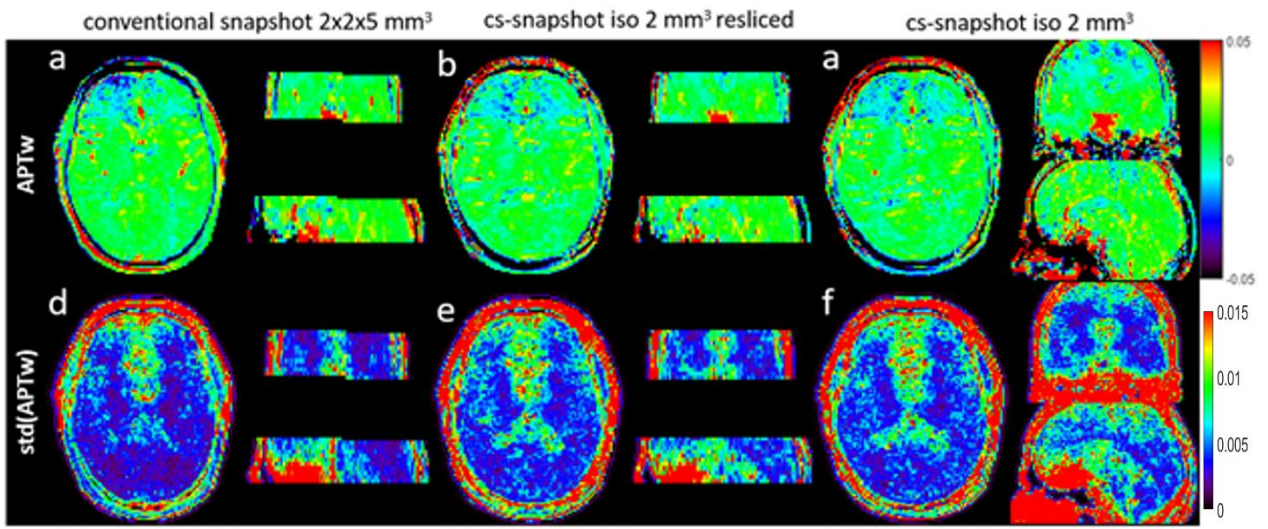

Figure 6: APTw maps (upper row) and standard deviation of a repeated measurement (lower row) of the established conventional snapshot CEST with limited brain coverage (a) compared to a 2 mm isotropic cs-snapshot CEST acquisition that was once resliced to match the resolution of the conventional protocol (b) and shown for full brain coverage in c).

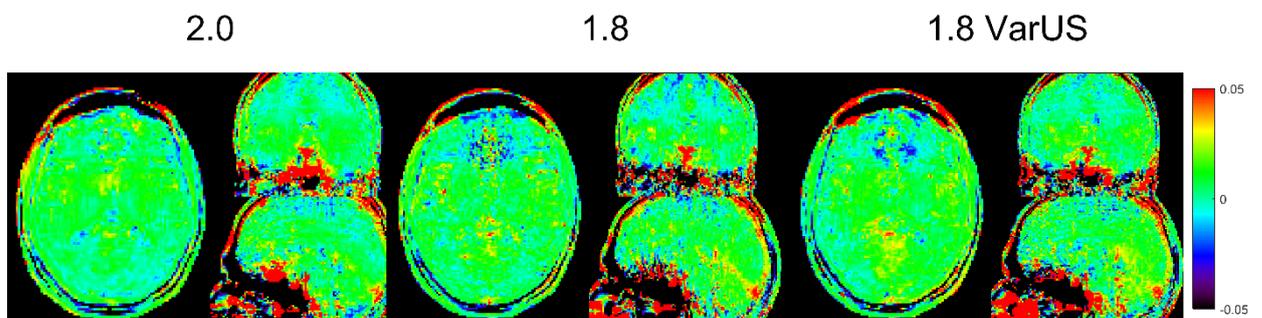

Figure 7: Comparison of different APTw cs-snapshot CEST maps for whole-brain acquisitions. Left: 2mm isotropic as in figure 6. Middle: 1.8 mm isotropic, Right: 1.8 mm isotropic with variable undersampling,

**Discussion**

A previous developed snapshot GRE CEST approach[2,14] was adapted for APTw imaging following the Pulseq-CEST standards[17] as defined in the APTw consensus paper[15]. Larger voxel sizes, slightly larger excitation flip angles, and presaturation with the high DC preparation (APTw_3T_001) yield APTw maps with decent image quality and the expected contrast in tumor tissue. With an optimized subsampling in frequency domain, we accelerate the snapshot APTw method to an acquisition time below 2 minutes. By additional undersampled acquisition and compressed sensing reconstruction this slab-selective approach could be extended to whole-brain coverage with isotropic resolution and without time penalty. It was shown that whole-brain 1.8mm isotropic snapshot CEST acquisitions are possible. Whole-brain acquisitions come with a higher acceleration factor and therefore an increase in standard deviation. However, it is not clear what standard deviation is sufficient in a clinical environment, where this increase in standard deviation might be readily traded off for a whole-brain coverage in combination with short scan times. Ultimately, the CEST APTw maps look promising, even for the 1.8 mm isotropic acquisition.

In all snapshot or multi-shot sequences the principal problem is that the prepared magnetization state decays during the readout. In spin-echo-based approaches, this decay is governed by $T_2$, in gradient-echo-based approaches this is governed by the Look-Locker time $T_{1LL}$. In gradient echo or gradient echo EPI it is governed by a combination of $T_{2*}$ and $T_{1LL}$. In the previous snapshot gradient-echo CEST article[2] it was derived that with approximately 3 times the time constant ($T_{1LL}$ in that case) a snapshot still yields a reasonable blurriness compared to the ideal image. The advantage of gradient-echo-based approaches is that $T_{1LL}$ is longer compared to $T_2$, thus more acquisitions can be performed for the same signal decay compared to a turbo-spin-echo sequence. This makes snapshot of a whole 3D volume possible by gradient-echo based approaches, while volumetric TSE approaches require typically 3 or more shots per pre-saturation offset which delays the acquisition process. This is why the snapshot approach with a measurement time of approximately 66 s beats existing multi-shot TSE-based APTw imaging sequences of 5 minutes duration and similar coverage[16]. When comparing the readout train time with GM T1 and T2, both TSE and GRE approaches have similar ratios, thus also the decay of the k-space signal and the resulting blurriness is comparable, but in different directions. Still, TSE with multiple shot lead also to intrinsically higher signal if centric reordered at the expense of scan time.

While snapshot approaches are still rare in MRI, as often steady-state approaches are favored, we think the benefits of snapshot approaches might even increase in the future. We showed herein that by using compressed sensing we can generate higher resolution and more coverage with the same number of k-space lines. Given the current progress in machine learning (ML)-based reconstruction[18–20], the resolution limitation of snapshot approaches further decreases, as even more image data can be generated with the same amount of acquired k-space data. Thus, the present article gives an outlook to the future and even more accelerated ML-snapshot CEST approaches.

Finally, we want to emphasize that the presented protocol is in agreement with the recent to recommendation of the APTw consensus white paper[7] regarding the APTw preparation, the offset sampling, as well as possible fat suppression pulses (see Supporting Figure 1 for comparison). Thus, we presented a standardized APTw imaging method that can be used in a clinical context, and in addition has accelerated high-resolution acquisition with whole brain coverage.

The presented GRE-based approach can be extended to other body regions, which will be investigated in the future. Future work will also encompass to use this approach with different preparation modules at 3T, as well as at 7T, where typically an even higher number of offsets needs to be acquired.

**Conclusion**

We presented (i) an optimized snapshot APTw protocol and (ii) a novel compressed-sensing snapshot CEST APTw sequence with whole brain coverage. Together this makes whole-brain APTw CEST acquisition times below 2 minutes possible in a clinical context.

**References**

1. van Zijl PCM, Yadav NN. Chemical exchange saturation transfer (CEST): What is in a name and what isn't?: CEST: What is in a Name and What Isn't? *Magn Reson Med*. 2011;65(4):927-948. doi:10.1002/mrm.22761

2. Zaiss M, Ehses P, Scheffler K. Snapshot-CEST: Optimizing spiral-centric-reordered gradient echo acquisition for fast and robust 3D CEST MRI at 9.4 T. *NMR in Biomedicine*. 2018;31(4):e3879. doi:10.1002/nbm.3879

3. Stehling MK, Turner R, Mansfield P. Echo-Planar Imaging: Magnetic Resonance Imaging in a Fraction of a Second. *Science*. 1991;254(5028):43-50. doi:10.1126/science.1925560

4. Akbey S, Ehses P, Stirnberg R, Zaiss M, Stöcker T. Whole-brain snapshot CEST imaging at 7 T using 3D-EPI. *Magn Reson Med*. 2019;82(5):1741-1752. doi:10.1002/mrm.27866

5. Lustig M, Donoho D, Pauly JM. Sparse MRI: The application of compressed sensing for rapid MR imaging. *Magn Reson Med*. 2007;58(6):1182-1195. doi:10.1002/mrm.21391

6. Herz K, Mueller S, Perlman O, Zaitsev M, Knutsson L, Sun PZ, Zhou J, Zijl P, Heinecke K, Schuenke P, Farrar CT, Schmidt M, Dörfler A, Scheffler K, Zaiss M. Pulseq-CEST: Towards multi-site multi-vendor compatibility and reproducibility of CEST experiments using an open-source sequence standard. *Magn Reson Med*. 2021;86(4):1845-1858. doi:10.1002/mrm.28825

7. Zhou J, Zaiss M, Knutsson L, Sun PZ, Ahn SS, Aime S, Bachert P, Blakeley JO, Cai K, Chappell MA, Chen M, Gochberg DF, Goerke S, Heo H, Jiang S, Jin T, Kim S, Laterra J, Paech D, Pagel MD, Park JE, Reddy R, Sakata A, Sartoretti-Schefer S, Sherry AD, Smith SA, Stanisz GJ, Sundgren PC, Togao O, Vandsburger M, Wen Z, Wu Y, Zhang Y, Zhu W, Zu Z, van Zijl PCM. Review and consensus recommendations on clinical APT -weighted imaging approaches at 3T : Application to brain tumors. *Magnetic Resonance in Med*. 2022;88(2):546-574. doi:10.1002/mrm.29241

8. Griswold MA, Jakob PM, Heidemann RM, Nittka M, Jellus V, Wang J, Kiefer B, Haase A. Generalized autocalibrating partially parallel acquisitions (GRAPPA). *Magn Reson Med*. 2002;47(6):1202-1210. doi:10.1002/mrm.10171

9. Rofsky NM, Lee VS, Laub G, Pollack MA, Krinsky GA, Thomasson D, Ambrosino MM, Weinreb JC. Abdominal MR Imaging with a Volumetric Interpolated Breath-hold Examination. *Radiology*. 1999;212(3):876-884. doi:10.1148/radiology.212.3.r99se34876

10. Lugauer F, Wetzl J, Forman C, Schneider M, Kiefer B, Hornegger J, Nickel D, Maier A. Single-breath-hold abdominal $T_{1}$ T 1 mapping using 3D Cartesian Look-Locker with spatiotemporal sparsity constraints. *Magn Reson Mater Phy*. 2018;31(3):399-414. doi:10.1007/s10334-017-0670-8

11. Uecker M, Lai P, Murphy MJ, Virtue P, Elad M, Pauly JM, Vasanawala SS, Lustig M. ESPIRiT-an eigenvalue approach to autocalibrating parallel MRI: Where SENSE meets GRAPPA. *Magn Reson Med*. 2014;71(3):990-1001. doi:10.1002/mrm.24751

12. Schuenke P, Windschuh J, Roeloffs V, Ladd ME, Bachert P, Zaiss M. Simultaneous mapping of water shift and B $_1$ (WASABI)—Application to field-Inhomogeneity correction of CEST MRI data. *Magn Reson Med*. 2017;77(2):571-580. doi:10.1002/mrm.26133

13. Jezzard P, Balaban RS. Correction for geometric distortion in echo planar images from B0 field variations. *Magn Reson Med*. 1995;34(1):65-73. doi:10.1002/mrm.1910340111

14. Deshmane A, Zaiss M, Lindig T, Herz K, Schuppert M, Gandhi C, Bender B, Ernemann U, Scheffler K. 3D gradient echo snapshot CEST MRI with low power saturation for human studies at 3T. *Magnetic Resonance in Medicine*. 2019;81(4):2412-2423. doi:10.1002/mrm.27569

15. Herz K, Lindig T, Deshmane A, Schittenhelm J, Skardelly M, Bender B, Ernemann U, Scheffler K, Zaiss M. T1ρ-based dynamic glucose-enhanced (DGEρ) MRI at 3 T: method development and early clinical experience in the human brain. *Magn Reson Med*. 2019;82(5):1832-1847. doi:10.1002/mrm.27857

16. Zhou J, Heo HY, Knutsson L, van Zijl PCM, Jiang S. APT-weighted MRI: Techniques, current neuro applications, and challenging issues. *Journal of Magnetic Resonance Imaging*. 2019;50(2):347-364. doi:10.1002/jmri.26645

17. Herz K. *Pulseq-Cest-Library*.; 2021. Accessed May 31, 2022. https://github.com/kherz/pulseq-cest-library

18. Knoll F, Murrell T, Sriram A, Yakubova N, Zbontar J, Rabbat M, Defazio A, Muckley MJ, Sodickson DK, Zitnick CL, Recht MP. Advancing machine learning for MR image reconstruction with an open competition: Overview of the 2019 fastMRI challenge. *Magn Reson Med*. 2020;84(6):3054-3070. doi:10.1002/mrm.28338

19. Hammernik K, Klatzer T, Kobler E, Recht MP, Sodickson DK, Pock T, Knoll F. Learning a variational network for reconstruction of accelerated MRI data: Learning a Variational Network for Reconstruction of Accelerated MRI Data. *Magn Reson Med*. 2018;79(6):3055-3071. doi:10.1002/mrm.26977

20. Wang G. A Perspective on Deep Imaging. *IEEE Access*. 2016;4:8914-8924. doi:10.1109/ACCESS.2016.2624938

# Supporting Info / Outlook

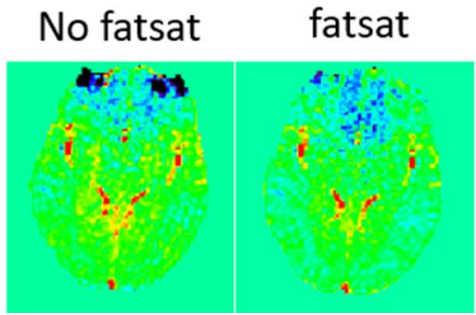

Supporting Figure S1: Necessity and possibility of fat saturation pulses. Fat artifacts visible in the frontal lobe (a) get efficiently removed by only single fat suppression pulse before the snapshot readout (b). This also improves the images of the tumor patient (c) in that are (same patient as Figure 3).

Finally, without changing the preparation or the readout any further, but simply choosing zero-filled reconstruction of the data, yields higher resolved smooth images with only minimal loss in details (Figure 5). The surprising performance of this interpolation is probably due to the T1-Look-Locker filter inherent to the snapshot-GRE readout, and the previously applied Hamming filter that works in synergy with the zero-filling.

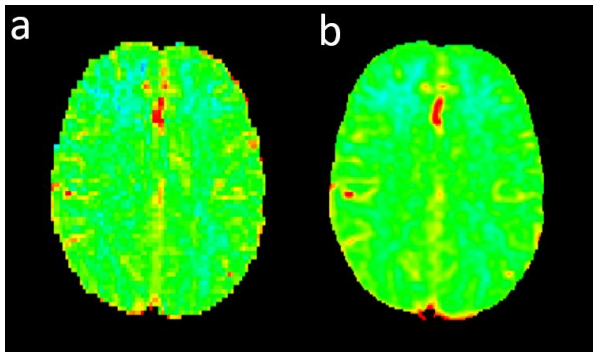

Supporting Figure S2: Possibility of two-fold in plane interpolation by filtered zero-filling